\documentclass[aps,pra,twocolumn,groupedaddress,showpacs,superscriptaddress]{revtex4-1}
\usepackage{graphicx,amsmath}
\usepackage{amssymb}

\begin{document}
\title{Phase transition of Frustrated Ising model via D-wave Quantum Annealing Machine}
\author{Hayun Park}
\affiliation{Department of Liberal Studies, Kangwon National University, Samcheok, 25913, Republic of Korea}
\author{Hunpyo Lee}
\affiliation{Department of Liberal Studies, Kangwon National University, Samcheok, 25913, Republic of Korea}
\email{Email: hplee@kangwon.ac.kr}
\date{\today}

\begin{abstract}
We study the frustrated Ising model on the two-dimensional $L \times L$ square lattice with ferromagnetic (FM) 
nearest-neighbor and antiferromagnetic diagonal-neighbor interactions using the D-wave quantum annealing machine (D-QAM) with 5000+ 
qubits composed on structure of the Pegasus graph. As the former Monte Carlo and mean field results, we find the FM to stripe order 
phase transition, through observations of the magnetization $M$, energy, magnetic susceptibility and structure factor. 
We also analyze probability which occurs any $M$ at a given interaction for many quantum annealing shots to estimate
the shape of objective function $f$. The only one value of $M$ with specific phase is observed in the regions far from phase 
transition for many quantum annealing shots, while several values of $M$ with different possibilities are appeared in the regimes of 
phase transition. We guess that $f$ in the regimes of phase transition retains the multi-modal structure with several local 
minimums, due to the strong degeneracies caused by frustrations. Finally, we discuss fail of the quantum annealing simulations, 
through analysis of the number of the chains, defined as the same variable with $N$-qubits, as a function of $L$. 
\end{abstract}

% 71.10.Fd    Lattice fermion models (Hubbard model, etc.)
% 71.27.+a    Strongly correlated electron systems; heavy fermions
% 71.30.+h    Metal-insulator transitions and other electronic transitions
\maketitle

\section{Introduction\label{Introduction}}

To search for an optimal solution from many possible candidates is one of the most important and challenging 
problems in the economics, computational science and machine learning communities. If many possible ones 
are expressed in a monotonous objective function $f$, it can be easily determined by minimizing 
(or maximizing) of $f$. On the other hand, most important and interesting problems appeared in computational science and machine 
learning calculation show the complex multi-modal structures in $f$ which processes several local minimums of plausible solutions. 
Therefore, various numerical approaches like a simulated annealing based on a stochastic method have been developing to avoid a 
trap of local minimum~\cite{Martin1970,Kirkpatrick1983}.

\begin{figure}[tbh]
\includegraphics[width=0.975\columnwidth]{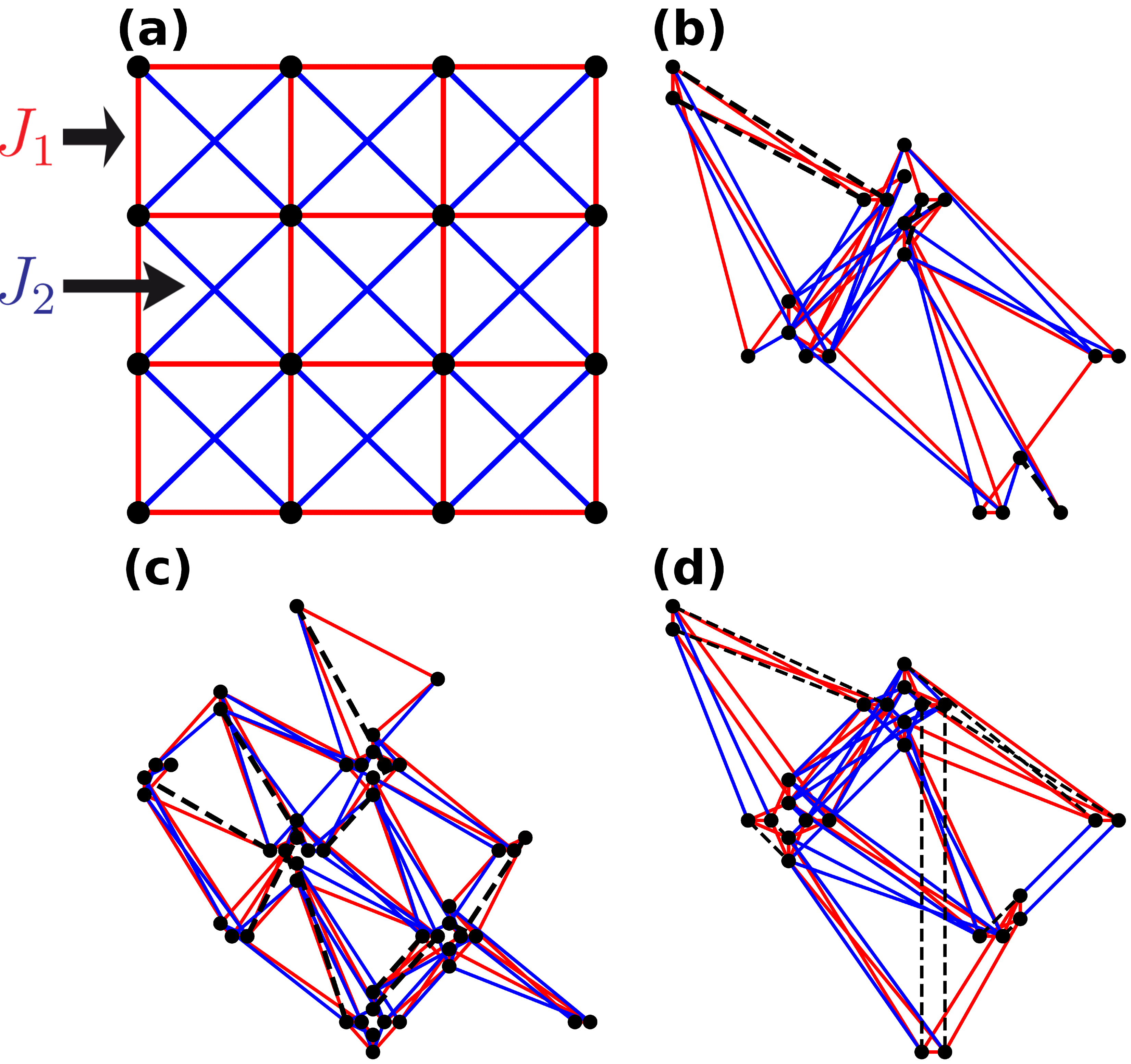}
\caption{\label{Fig1} (a) Frustrated lattice on two-dimensional $4 \times 4$ square lattice with nearest-neighbor and diagonal-
neighbor interactions of ferromagnetic (FM) order with $J_1>0$ and antiferromagnetic order with $J_2<0$, respectively. (b) The 
embedding with 16-sites on Perasus graph which topologically consists with (a) lattice of open boundary condition (OBC). (c) 
The embedding with 36-sites of OBC. (d) The embedding with 16-sites on Perasus graph which 
topologically consists with (a) lattice of periodic boundary condition. Here, the chains expressed as the same variables with 
$N$-qubits on the Pegasus graph are marked as the dotted line.}
\label{fig1}
\end{figure}

Recently, another attempt to solve optimization problems has been performing on the D-wave quantum annealing machine 
(D-QAM) based on superconducting qubits~\cite{Johnson2011}. A quantum annealing (QA) simulation is implemented on parameterized 
Hamiltonian of a transverse-field Ising model with binary quadratic form (BQF) modified from $f$ of the original optimization 
problem~\cite{Johnson2011,Kadowaki1998}. It has not only described the unconventional phases in physical systems, but also have 
applied for various engineering applications such as green energy production and traffic signal 
optimization~\cite{Inoue2021,Kairys2020,King2021,Irie2021,Teplukhin2021,Sharabiani2021}. On the other hand, more examinations and 
calculations on the D-QAM would be still required and highly beneficial, because they can provide more computation information and 
technique for an efficient utilization and application of D-QAM. In addition, there have been a few investigations via the D-QAM.

A frustrated Ising model on the two-dimensional (2D) square lattice at zero temperature $T$ is not only one of the good examples 
with several local traps in $f$, but also is an exactly unsolved and interesting system, which exhibits the exotic phases and phase 
transitions by competition among various orderings, contrary to the case of the Ising model with only nearest-neighbor 
interaction, where the exact result was known~\cite{Onsager1944}. Furthermore, the standard Monte Carlo (MC) simulations have not 
been well performed at zero temperature $T$, because the values of the Boltzmann weight function $e^{-\Delta E/T}$ for acceptances 
of MC step are converged to zero ($e^{-\infty}=0$) with decreasing $T$. Therefore, we believe that the frustrated Ising model would 
be a good applicant for analysis of the simulation results and techniques on the D-QAM.

In this paper, using the D-QAM with 5000+ qubits composed on structure of Pegasus graph in the amazon web services (AWS) we explore 
the frustrated Ising model on 2D $L \times L$ square lattice with competitions of between nearest-neighbor of ferromagnetic (FM) 
order and diagonal-neighbor of antiferromagnetic (AF) order, where the results were well known by the MC and mean field 
methods~\cite{Jin2012,Jin2013}. The Hamiltonian is given as
\begin{equation}\label{Eq1}
H = -J_1 \sum_{<i,j>} \sigma_i^z \sigma_j^z + J_2 \sum_{<<i,j>>} \sigma_i^z \sigma_j^z,
\end{equation}
where nearest- and diagonal-neighbors are denoted by $<i,j>$ and $<<i,j>>$, respectively. Here, we employ $J_1>0$ and $J_2>0$ which 
mean the preferences of FM and AF orderings, respectively. We confirm the FM to stripe order (SO) phase transition around $J_2/J_1 
\approx 0.5$ from results of the magnetization $M$, energy per site $E$,  magnetic susceptibility $\chi$ and structure factor 
$S(\vec{q})$ for $L=20$, through the QA simulation. These results consist with the former MC and mean field 
ones~\cite{Jin2012,Jin2013}. We also analyze probability, which presents any value of $M$ at a given $J_2/J_1$ for many QA shots, 
to predict the shape of $f$. Finally, we discuss limitations of the D-QAM by break of the chains, via observation of the 
number of the chains with meaning of the same variable with $N$-qubits as a function of $L$.

The paper is organized as follows: Section~\ref{Computation} gives a detailed description of the QA process on the D-QAM with 5000+ 
qubits. In Section~\ref{Result}, we discuss various results on the frustrated 2D square lattice Ising model at $T=0$
and limitation of the D-QAM by chops of the chains with different $N$-qubits. Finally, we present the conclusions in 
Section~\ref{Conclusion}.

\section{Annealing Method\label{Computation}}

For our QA simulation we use the D-QAM with 5000+ qubits composed on structure of Pegasus graph in the AWS. 
The QA is performed on the transverse field Ising model with BQF based on superconducting qubits~\cite{Johnson2011,Kadowaki1998}. 
Therefore, $f$ needs to be converted into the BQF in the first QA step. On the other hand, this transformation process is not 
necessary in our QA simulation, because $f$ was already expressed as BQF of Ising spins.

\begin{figure}[tbh]
\includegraphics[width=0.975\columnwidth]{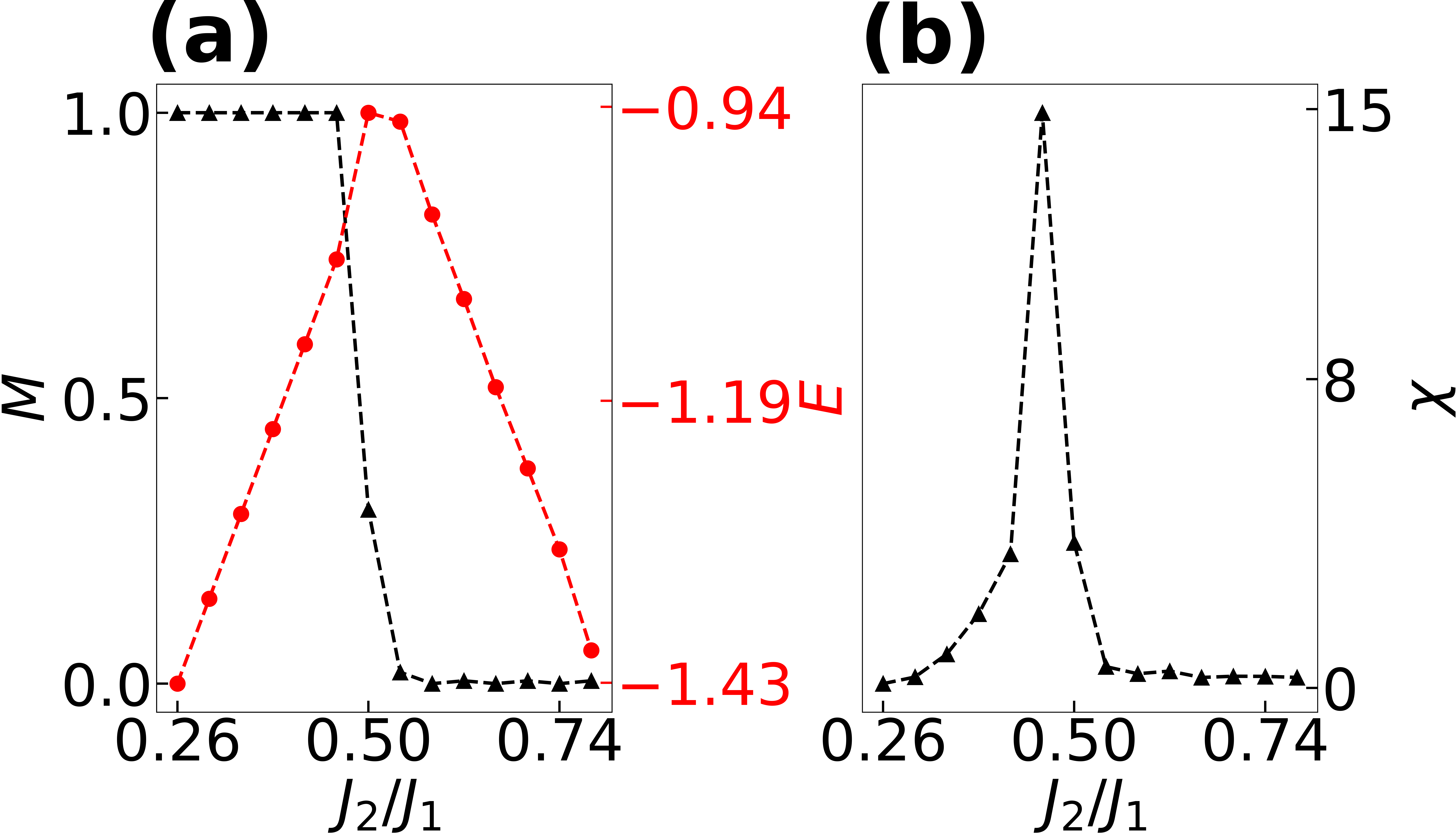}
\caption {\label{Fig2} (Color online) (a) (Left) Magnetization $M$ and (Right) energy per site $E$ as a function of $J_2/J_1$. (b) 
Magnetic susceptibility $\chi$ as a function of $J_2/J_1$. Quantum annealing (QA) simulations are performed on two-dimensional $L 
\times L$ square lattice with $L=20$.}
\end{figure}

Next QA step is to map the 2D $L \times L$ square lattice into the architecture of the Pegasus graph on the D-QAM with 5000+ 
qubits. Figs.~\ref{Fig1}(a) and (b) show an example of $4 \times 4$ square lattice with (red) nearest- and (blue) diagonal-neighbors 
interactions and an embedding with 16-site spins on the Pegasus graph with open boundary condition (OBC), respectively. Here, the 
chains with a meaning of the same variables should essentially introduce in the embedding, because the original lattice is 
topologically in discord with the architecture of Pegasus graph. Note that the chains are composed as $N$-qubits, where $N$ is the 
number of qubits. The chains on the Pegasus graph are marked as the dotted line in Figs.~\ref{Fig1} (b), (c) and (d). As the results 
shown in Figs.~\ref{Fig1} (b) with $L=4$ and (c) with $L=6$ in OBC, the number of the chains and $N$ are increasing with increasing 
the size of $L$ in the original lattice. In addition, the number of the chains and $N$ in periodic boundary condition (PBC) are 
more than those in OBC, respectively, as results displayed in Figs.~\ref{Fig1} (b) with OBC and (d) with PBC. The qualities of 
the QA calculations mainly suffer with increasing the number of the chains and $N$, due to high possibilities of a break of chain. 
Therefore, to find the embedding with the smallest number of the chains in many possibilities of embeddings is an important process 
for qualities of the QA simulations. We employ the iterative embedding algorithm offered in the D-QAM with 5000+ qubits of the 
AWS~\cite{Cai2014,Dattani2019}. 

\begin{figure*}[tbh]
\includegraphics[width=0.925\textwidth]{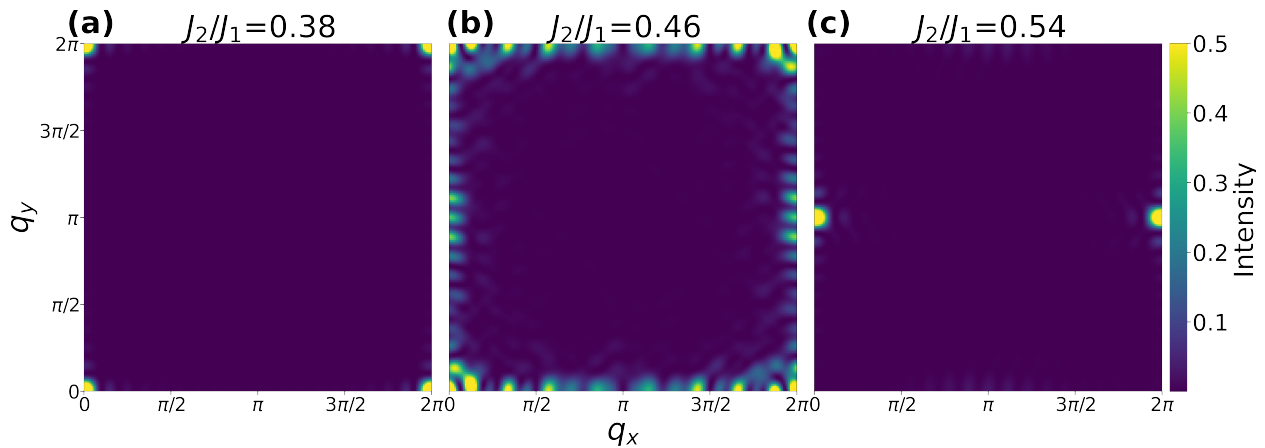}
\caption {\label{Fig3} (Color online) (Left) Structure factors for (a) $J_2/J_1=0.38$, (b) $0.46$ and (c) $0.54$. FM, 
coexistance and stripe order (SO) phases are observed in (a), (b) and (c), respectively.}
\end{figure*}

Finally, the QA is performing on BQF which is defined as 
\begin{equation}
\text{BQF} = f + \sum_{i=1}^{L^2} g(i),
\end{equation}
where $g(i)$ is a constraint for the optimal solution. The detailed procedures of the QA process are following. (i) 
Initially, we randomly select $g(i)$ to find the specious solution in many shots. (ii) After determining the specious solution, we 
adjust $g(i)$ to search for more optimal solution, through consulting the former configuration and constraint. (iii) We perform the 
QA with modified $g(i)$ again and find the specious solution which is closer to optimal solution. (iv) In the end the 
total energy is converged into the lowest one with the optimal solution, after doing several (ii) and (iii) processes. We perform 
the QA shots of 10000 times to find the configurations with the lowest energy in final $g(i)$, and obtain all measured quantities 
from the configurations with the lowest energy.

\section{Result\label{Result}}

\begin{figure}[tbh]
\includegraphics[width=0.98\columnwidth]{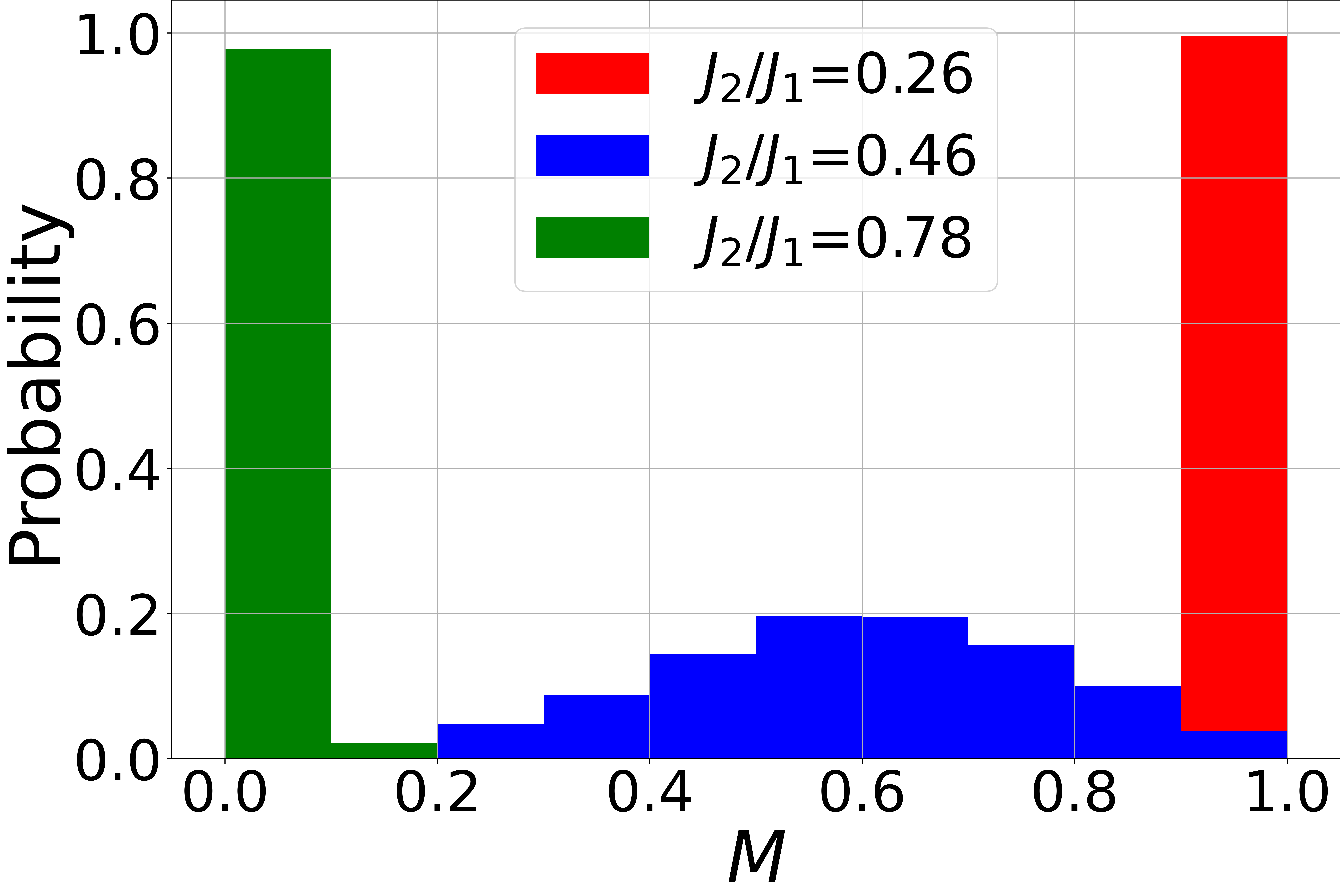}
\caption {\label{Fig4} (Color online) Probability where any $M$ is observed for the QA shots of 10000 times for $J_2/J_1=0.26$, 
$0.46$ and $0.78$. The possibilities indicate only one value of $M$ in the regimes with PM and SO phases of $J_2/J_1=0.26$ and 
$0.78$ respectively, while those spread out as several $M$ around phase transition with $J_2/J_1=0.46$.}
\end{figure}

We use the 2D $L \times L$ square lattice with $L=20$ for the results shown in Fig.~\ref{Fig2}, Fig.~\ref{Fig3} and Fig.~\ref{Fig4}. 
We first plot $M$ and $E$ as a function of $J_2/J_1$ in Fig.~\ref{Fig2}(a). The FM phase with spin alignments is clearly 
observed below $J_2/J_1=0.46$ in the results of $M$. The phase transition, where $M$ rapidly drops, is shown at 
$J_2/J_1 \approx 0.5$. $M$ are completely disappeared above $J_2/J_1=0.56$. $E$ displays the highest energy 
around phase transition with $J_2/J_1 \approx 0.5$. We also present $\chi$ expressed as $\chi = \vert <M^2> - <M>^2 
\vert$ in Fig.~\ref{Fig2}(b). We confirm that the magnetic fluctuations are rapidly increasing around phase transition, because the 
FM and AF orderings are strongly competing.  

Next we present $S(\vec{q})$ with different $J_2/J_1$ in Figs.~\ref{Fig3}(a)-(c). $S(\vec{q})$ is defined as 
\begin{equation}
S(\vec{q}) = \sum_{i,j} <\sigma_i^z \sigma_j^z> e^{i\vec{q}\cdot(\vec{R_i} - \vec{R_j})},
\end{equation}
where $\vec{R_{i,j}}$ is the relative position of two spins and $\vec{q}$ is the wave vector in reciprocal space. It 
not only supplies information of the spatial configurations in the real space, but also can be directly compared to the results 
measured by the neutron diffuse scattering spectrum. The strong intensity is clearly seen at $S(\vec{q}=(0,0))$ with FM phase in 
Fig.~\ref{Fig3}(a) with $J_2/J_1=0.38$. It is disappeared at $S(\vec{q}=(0,0))$ with increasing $J_2/J_1$ 
and spread out from $\vec{q}=(0,0)$ to $(\pi,0)$ in Fig.~\ref{Fig3}(b) with $J_2/J_1=0.46$. These results mean the 
coexistance phase around phase transition. In the end it is only observed at $(\pi,0)$ with SO phase in Fig.~\ref{Fig3}(c) with 
$J_2/J_1=0.56$ of the strong diagonal AF interaction.

In the following we analyze probability, where any value of $M$ is observed for the QA shots of 10000 times, to predict the shape of 
$f$ for $J_2/J_1=0.26$, $0.46$ and $0.78$. The only one value of $M$ with particular phase is clearly seen at $J_2/J_1=0.26$ and 
$0.78$ with FM and SO phases, respectively in Fig~\ref{Fig4}. We guess that the shape of $f$ in these 
regions would have monotonous structure with a deep global minimum. On the other hand, several values of $M$ are appeared with 
different possibilities in the regimes of phase transition with $J_2/J_1=0.46$ in Fig~\ref{Fig4}. We think that the strong 
degeneracies appeared by frustrations would make the complex multi-modal structure with plausible solutions in $f$. Therefore, we 
think that many QA shots would be trapped at the local minimums with different possibilities.

\begin{figure}
\includegraphics[width=0.98\columnwidth]{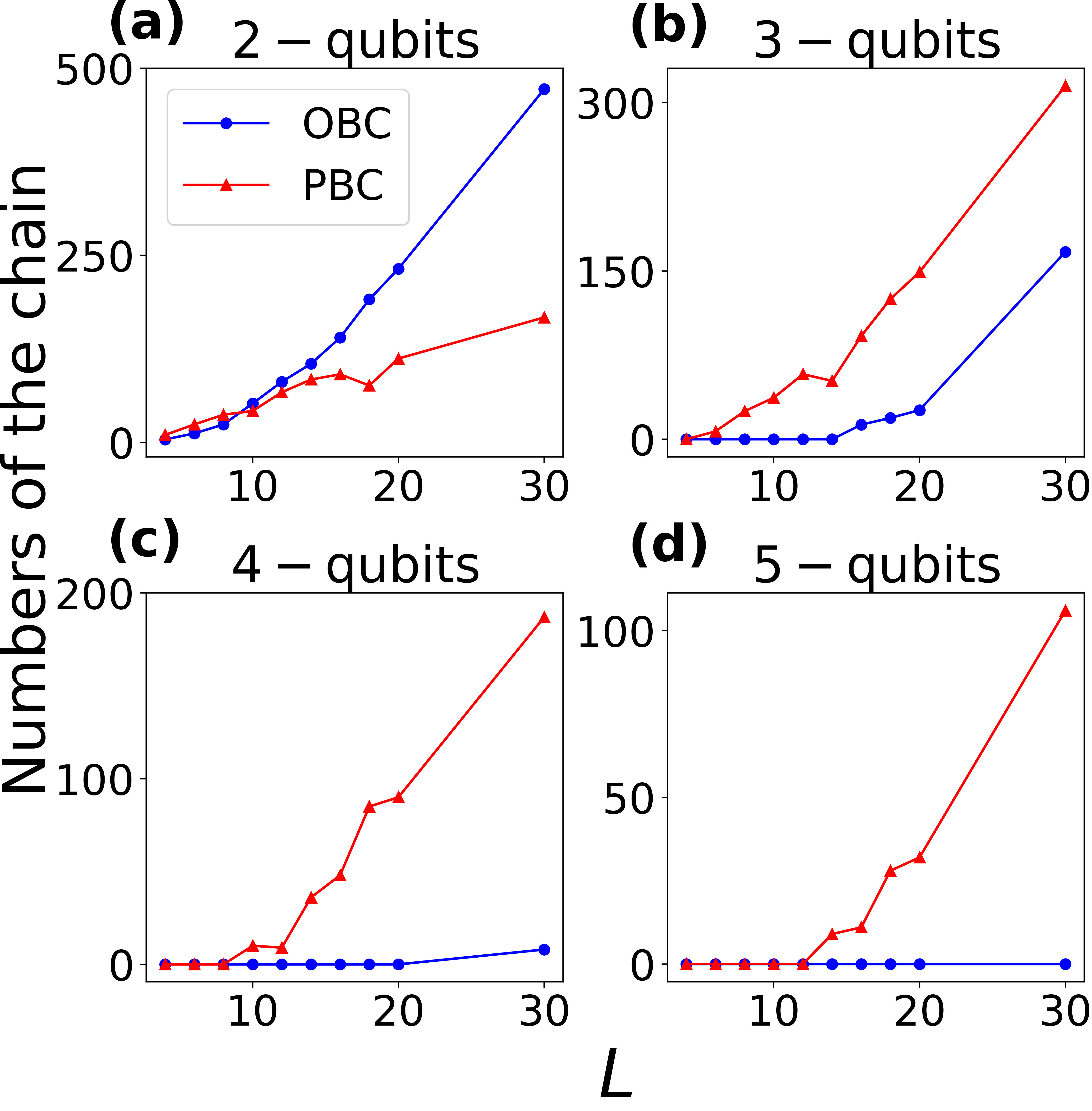}
\caption {\label{Fig5} (Color online) The numbers of the chain as a function of $L$ in both OBC and PBC for (a) $2$-qubits, 
(b) $3$-qubits, (c) $4$-qubits and (d) $5$-qubits.}
\end{figure}

Finally, we discuss fail of the QA simulations happened by breaks of the chains, through measurement of the number of the 
chains with several $N$-qubits as a function $L$. The results are shown in Figs.~\ref{Fig5}(a)-(d). The numbers of the chains in 
both OBC and PBC are commonly increasing with increasing $L$ in the chains composed of $2$-qubits in Fig.~\ref{Fig5}(a). The numbers 
of the chain in PBC are rapidly increasing with increasing $L$ in the chains with $N$-qubits, while they in OBC are not largely 
increasing till $L=30$ in Figs.~\ref{Fig5}(b)-(d). We confirm that the QA simulations have been well performed on the sizes smaller 
than $L=30$ in OBC, while they have not been only worked on the sizes smaller than $L=16$ in PBC. In the end we stress that 
$2$-qubits in the chains are not well broken, while $N$-qubits more than $2$-qubits in those are easily separated.

\section{Conclusion\label{Conclusion}}
Via the D-QAM with 5000+ qubits composed on structure of Pegasus graph in the AWS, we investigate the frustrated Ising 
model on the 2D $L \times L$ square lattice at $T=0$, where frustrations are present by AF diagonal-neighbor interaction. We 
confirm FM to SO phase transition, which consists with the former MC and mean field results, through measurements of the 
$M$, $E$, $\chi$ and $S(\vec{q})$. We also analyze probability which happens the particular $M$ at a given $J_2/J_1$ for many QA 
shots. The possibilities indicate only one $M$ in the regimes with FM and SO phases far from phase transition. We 
think that the deep global minimum might be observed in $f$ of these regions. On the other hand, those spread out as several $M$ 
around phase transition. Owing to the strong degeneracies caused by frustrations, we guess that several local traps with 
plausible solutions in $f$ might extensively appear around phase transition. Finally, we discuss limitation of the D-QAM 
with 5000+ qubits, through analysis of the number of the chains, which are defined as the same variables with $N$-qubits on 
architecture of the Pegasus graph, as a function of $L$.

We would like to note that the QA simulations on D-QAM are limited as the 2D lattice with small $L$ because of breaks of the chains, 
unlike the MC and mean field approaches which can consider one with large $L$~\cite{Jin2012,Jin2013}. Therefore, we have not 
discussed the type of phase transition and the critical exponents, appeared in the former MC and mean field results, in our 
manuscript~\cite{Jin2012,Jin2013}.    

\section{Acknowledgements}
We would like to thank Min-Chul Cha, Minho Kim and Myeonghun Park for useful discussions. This work was supported by Ministry of 
Science through NRF-2021R1111A2057259.

\end{document}